\documentstyle[preprint,aps]{revtex}

\begin{document}
\draft
\title{Interband mixing between two-dimensional states localized in a surface
quantum well and heavy hole states of the valence band in narrow gap
semiconductor}
\author{V.A.Larionova and A.V.Germanenko}
\address{Institute of Physics and Applied 
Mathematics, \\ Ural 
University, Ekaterinburg 620083, Russia}
\date{\today}
\maketitle

\begin{abstract}
Theoretical calculations in the framework of Kane model have been carried
out in order to elucidate the role of interband mixing in forming the energy
spectrum of two-dimensional carriers, localized in a surface quantum well in
narrow gap semiconductor. Of interest was the mixing between the 2D states
and heavy hole states in the volume of semiconductor. It has been shown that
the interband mixing results in two effects: the broadening of 2D energy
levels and their shift, which are mostly pronounced for semiconductors with
high doping level. The interband mixing has been found to influence mostly
the effective mass of 2D carriers for large their concentration, whereas it
slightly changes the subband distribution in a wide concentration range.
\end{abstract}

\narrowtext

\section{Introduction}

\label{sec:intro}

Interband mixing effects attract much interest because they can play a
decisive part in forming the energy spectrum of low-dimensional systems.
Such is in the case for InAs-GaSb semimetallic superlattices, whose physical
properties are determined by the mixing between valence band states of GaSb
and conduction band states of InAs at the interfaces. \cite
{Altarelli,Fassolino,Sakaki}

Analogous situation takes place in a metal--insulator--semiconductor
structure based on a narrow gap semiconductor. \cite
{Sobkowicz,Nachev,Ziegler,Freytag,Kunze,Kunze1} A small value of the energy
gap in the semiconductor leads to the fact that at energies below the top of
the valence band the two-dimensional (2D) states localized in a surface
quantum well can mix with the valence band states and 2D carriers can tunnel
from the space charge region into the valence band states in the volume of
semiconductor. This results in a shift
of the 2D subband energies and a broadening of these levels. \cite{Sobkowicz,Nachev,Ziegler,Freytag} Due to
high value of the effective mass, the probability of tunneling of 2D
carriers into the heavy hole states is small. Therefore the interband mixing with
the heavy hole states was neglected. This approach is justified for
semiconductors with low impurity density, because the surface potential in
such materials is smooth in the shape, and the 2D states localized in the
quantum well are spatially separated from the valence band by a wide region
of forbidden energies.

In semiconductors with high doping level the effects of interband tunneling
of 2D electrons into the heavy hole states can be stronger. The influence of
these effects on the energy spectrum and broadening of the 2D states
localized in the surface quantum well in the narrow gap semiconductor
(HgCd)Te is discussed in the present article.

\section{Theoretical model}

\label{sec:theor}

A small value of the energy gap ($E_g$) in semiconductor investigated makes
it necessary to employ a multiband Hamiltonian in the $kP$-based
calculations of the energy spectrum. We started from the multiband
Hamiltonian derived from Kane's model making the usual assumption that the
energy difference between the valence $\Gamma_8$-band and spin-orbit
split-off $\Gamma_7$-band is infinite. To consider the effects involving the
heavy hole states, the interaction with remote bands has been taken into
account by a standard procedure of including additional terms with $\gamma$%
-parameters in the Hamiltonian. \cite{Bir-Pikus} An isotropic approximation
has been used.

The Kane Hamiltonian, which is a $6\times 6$ matrix in the above
assumptions, can be block diagonalized, when the axis $z$ is chosen to be
normal to the interface and the axis $y$ along the direction of the carrier
motion. In this case the Hamiltonian is two $3\times 3$ matrices for two
groups of states. One of the matrices is defined by \widetext
\begin{equation}
\hat{H}^+= \left( 
\begin{array}{ccc}
E^{\Gamma_6}(z) & i\sqrt{\frac{2}{3}}P\left(\frac{k}{2}- \frac{\partial}{%
\partial z}\right) & \frac{i}{\sqrt{2}}Pk \\ 
i\sqrt{\frac{2}{3}}P\left(-\frac{k}{2}- \frac{\partial}{\partial z}\right) & 
{\cal G} & -\sqrt{3}\frac{\hbar^2}{2m_0}\gamma \left(k^2+2k\frac{\partial}{%
\partial z}\right) \\ 
-\frac{i}{\sqrt{2}}Pk & -\sqrt{3}\frac{\hbar^2}{2m_0}\gamma \left(k^2- 2k%
\frac{\partial}{\partial z}\right) & {\cal F}
\end{array}
\right),  \label{eq1}
\end{equation}
\[
{\cal G}=E^{\Gamma_8}(z)+\frac{\hbar^2}{2m_0}\left((\gamma_1+2\gamma) \frac{%
\partial^2}{\partial z^2} -(\gamma_1-\gamma)k^2\right)+e\varphi(z), 
\]
\[
{\cal F}=E^{\Gamma_8}(z)+\frac{\hbar^2}{2m_0}\left((\gamma_1- 2\gamma) \frac{%
\partial^2}{\partial z^2} - (\gamma_1+\gamma)k^2\right)+e\varphi(z), 
\]
\narrowtext

where $k\equiv k_y$, $P$ is the momentum matrix element, $\gamma_1$, $\gamma$
are the parameters, which describe the interaction with remote bands, $%
\varphi (z)$ is the electrostatic potential, and $E^{\Gamma_6}(z)$, $%
E^{\Gamma_8}(z)$ are the energies of the conduction and valence band edges,
respectively. Here, the assumption has been made that the values of $P$, $%
\gamma_1$, and $\gamma$ are independent of coordinate.

The second matrix $H^-$ is obtained from $H^+$ by replacing $k$ by $-k$.

The electrostatic potential can be derived from the Poisson equation. We
assumed $\varphi (z)$ to be parabolic in the space charge region and
constant in the volume of semiconductor: 
\begin{equation}
\varphi(z)= \left\{ 
\begin{array}{ll}
\varphi_s(1-z/L)^2, & 0\leq z\leq L \\ 
0, & z>L,
\end{array}
\right.  \label{eq2}
\end{equation}
where 
\[
\begin{array}{cc}
\varphi_s=\varphi(0), & L=\left(\frac{2\kappa\kappa_0\varphi_s}{e(N_A-N_D)}%
\right)^{1/2}
\end{array}
. 
\]
Here, $N_A-N_D$, and $\kappa$ stand for density of acceptors uncompensated,
and dielectric constant, respectively. Expression (\ref{eq2}) is a good
approximation for the case when the concentration of 2D electrons is less
than the value of $(N_A-N_D)L$.

Thus, Schr\"odinger equation is two independent systems of differential
equations of the second order. For solving these systems boundary conditions
have to be deduced. The introduction of the boundary conditions is based on
a model proposed in Refs. \onlinecite{Sobkowicz} and \onlinecite{our} for
calculation of the energy spectrum of 2D states in MIS structures, based on ordinary and
inverted semiconductors, respectively. As in these papers, we supposed (see
Fig.\ref{fig1}a): (i) the band structure of insulator to be identical to
that of semiconductor; (ii) the energy gap in the insulator to be much
greater than that in semiconductor, and the conduction and valence band
offsets, denoted as $D_c$ and $D_v$, respectively, to exceed the value of $%
E_g$; (iii) the energy bands in the insulator to be ``flat", i.e. $%
\varphi(z<-d)=\text{const}$, and (iv) the structure to have a smooth
insulator/semiconductor interface of width $d$ with linear dependence of the
band-edge energies on $z$ coordinate.\cite{our}

Since a general solution of the Kane Hamiltonian for $\varphi (z)=\text{const%
}$ is known, the wave function on the insulator ($z<-d$) and semiconductor ($%
z>L$) sides of the structure can be simply derived.

For $z<-d$ it contains only the terms exponentially vanishing deep into the
insulator.
To choose the boundary conditions on the semiconductor side let us briefly
consider three different cases, which can be realized depending on the
energy and longitudinal component of quasimomentum of 2D state. First, the
2D state is not degenerate with the valence band (region I in Fig.\ref{fig1}%
b). In this case the wave function exponentially diminishes deep into the
volume of semiconductor. Second, the 2D state is degenerate only with the
heavy hole valence band (region II). In this case the normal component of
the light hole quasimomentum is imaginary. Therefore, the wave function in
this case is a superposition of the light hole term, which diminishes
exponentially deep into the semiconductor, and two oscillating terms
presenting the plane wave associated with the heavy hole. Third, the 2D
state is in resonance with both heavy and light hole valence bands (region
III) and the wave function is a superposition of two plane waves
corresponding to the light and heavy holes. In the present paper we focus
our attention on the second case.

Thus, starting from the exact solution of the Kane Hamiltonian on the
insulator side we numerically integrated the differential equation systems
through the space charge region and chose those solutions which satisfy the
boundary conditions on the semiconductor side.

It is clear that in region II the eigenvalue problem has a continuous
spectrum of solutions, i.e. at any energy value we can find the wave
function satisfying the boundary conditions. To define the energy level
associated with the resonant 2D states and its broadening caused by the
interband mixing between the 2D states and heavy hole states, the Levinson's
theorem has been applied. \cite{Sobkowicz,Freytag}

\section{Discussion}

\label{sec:disc}

To analyze peculiarities of the energy spectrum of 2D states arising from
the interband mixing, we have performed numerical calculations using
realistic parameters which are close to the parameters of the sample
investigated in Ref.\onlinecite{ng}. They are the following: $E_g=50$ meV, $%
N_A-N_D=6\times 10^{17}\text{ cm}^{-3}$, $\kappa =20$. The calculations have
been carried out using two sets of $\gamma $-parameters: $\gamma _1=2.0$, $%
\gamma =0$ (Ref.\onlinecite{our}) and $\gamma _1=4.5$, $\gamma =1.0$ (Ref.%
\onlinecite{Guldner}). To make identical the dispersion law of bulk electron
states calculated with different sets of $\gamma $-parameters, we changed
slightly the value of the momentum matrix element $P$ from $8.0\times
10^{-8} $ eV$\cdot $cm, for the first set, to $8.1\times 10^{-8}$ eV$\cdot $%
cm, for the second one. Parameters of the insulator/semiconductor interface
are the same as in Ref. \onlinecite{our}: $D_c=2$ eV, $D_v=1$ eV, and $d=10$
\AA .

Figure \ref{fig2} shows the energy spectrum of 2D states calculated with
different sets of $\gamma$-parameters as compared with the energy spectrum
of 2D states calculated in the infinite heavy hole mass approximation with a
so-called ``mid-gap" boundary condition on the semiconductor side. \cite
{Marques} Two branches of the energy spectrum specified by $k^+$ and $k^-$
are the results of calculation for two groups of states. These branches are
two ``spin" branches of the ground 2D subband split by spin-orbit
interaction in an asymmetric quantum well. \cite
{ng,Bychkov,Ohkawa,Wollrab,Sizmann} It is clearly seen from the figure that
the interband mixing of 2D states with the heavy hole states results in the
energy shift of 2D subbands. The smaller is the heavy hole effective mass ($%
m_h=m_{0}/(\gamma_1-2\gamma)$) the lower is the energy position of the 2D
subbands.

One more important feature of the energy spectrum of 2D states appearing due
to interband mixing is the broadening of 2D energy levels. This broadening
is shown in Fig.\ref{fig2} as hatched region. As is seen the broadening of
2D levels is negligibly small for energies close to the top of the valence
band in the volume of semiconductor. At more negative energies both 2D
``spin" sublevels become broad. The value of broadening is different for $k^+
$ and $k^-$ sublevels for a fixed energy. It increases with decreasing
energy or quasimomentum value. The maximum value of the broadening is about $%
10$ meV. It is comparable to the broadening of 2D states appearing due to
tunneling into the light hole states which has been calculated in Ref. %
\onlinecite{Nachev} for a sample with close parameters. On further
decreasing quasimomentum the 2D states fall in resonance with the light hole
valence band (region III in Fig. \ref{fig1}b). Note that the broadening
value slightly depends on the heavy hole effective mass.

Thus, the interband mixing between the 2D states and heavy hole states leads
to two effects: the shift of subband energies and the broadening of 2D levels.
Let us consider how these effects can manifest themselves in experiment.

Traditional experimental methods, such as galvanomagnetic investigations,
volt- capacitance spectroscopy, give information about the carriers at the
Fermi energy. In $p$-type (HgCd)Te the Fermi level lies near the top of the
valence band in the energy range $-5$...$+5$ meV at low temperatures. As was
shown above for these energies the broadening of 2D states is small and
therefore these methods must not be sensitive to this effect. Tunneling
investigations allow to get information about the 2D states in a wide energy
range including the energies, where the broadening is large. Besides, the
maximum value of broadening is comparable with the cyclotron energy in
magnetic fields right up to $2$ T. This can result in additional broadening
of resonance peaks in magneto-optical and cyclotron resonance experiments.

As to the effect connected with the influence of interband mixing on the
dispersion law of 2D states traditional experimental methods give no way of
direct deducing a quasimomentum dependence of the energy of 2D states for a
fixed value of a surface quantum well. For example, in tunneling experiments 
\cite{our,rus} a voltage bias, applied to the structure, changes not only
the energy tested, but surface potential as well. The value of the surface
potential is usually used as a fitting parameter in analyzing experimental
data, for example, dependences of 2D subband concentration and the effective
mass of 2D carriers on their total concentration. Figure \ref{fig3} shows
the relationship between 2D subband concentration and the total
concentration of 2D electrons, calculated using two sets of $\gamma$%
-parameters and infinite heavy hole effective mass. The concentration of 2D
electrons is changed by varying the surface quantum well depth from $250$ to 
$320$ mV. It is clearly seen from the figure that the interband mixing does
not cause essential changes in the subband distribution of 2D electrons. One
can see only slight increase in ratio between concentrations of 2D electrons
in $k^+$ and $k^-$ ``spin" subbands with decreasing the heavy hole
effective mass. Thus, in spite of the fact that the interband mixing results
in shifting the energy positions of 2D subbands, it does not strongly
influence the subband distribution of 2D electrons.

The effective mass of carriers is a differential characteristic of the
energy spectrum and it might be more sensitive to the interband mixing
effects. We have calculated the effective mass of 2D electrons $m^{*}=\hbar
^2k\left( \partial E/\partial k\right) ^{-1}$ at $E=E_F$ as a function of
their total concentration (Fig. \ref{fig4}). As is clearly seen for large
concentration of 2D carriers ($N_S>0.55\times 10^{12}\text{cm}^{-2}$) the
effective mass of 2D electrons, calculated in the model taking into account
the interband mixing (solid and dashed curves), is larger than that
calculated with the infinite heavy hole mass (dotted curves). The difference
increases with increasing the total concentration of 2D electrons. It is
evident that the concentration dependence of the effective mass, calculated
with $\gamma _1=4.5$, $\gamma =1.0$ deviates mostly from the results
obtained in the infinite heavy hole mass approximation. For $k^{+}$-``spin''
states this difference is larger and at $N_S\simeq 10^{12}\text{cm}^{-2}$ it
comes to more than $10\%$. 

A distinctive situation occurs for small concentrations of 2D electrons ($%
N_S<0.45\times 10^{12}\text{cm}^{-2}$) for $k^{+}$-group of states. The
effective mass, calculated with $\gamma $-parameters, monotonically
decreases with decreasing the total concentration, whereas the effective
mass, calculated in the framework of the infinite heavy hole mass
approximation, goes through a minimum, and at some value of the total
concentration it becomes larger than the effective mass calculated with $%
\gamma $-parameters. To explain this peculiarity, let us turn to Figure \ref
{fig2}. One can see that the bottom of $k^{+}$-subband lies at $k\neq 0$
(the dotted curve in the figure). As a consequence, the effective mass $%
m^{*}=\hbar ^2k\left( \partial E/\partial k\right) ^{-1}$ turns to infinity
on the bottom and is sufficiently large in its vicinity. For the low
concentrations of 2D electrons, this portion of the energy spectrum is found
to be near the Fermi level for the case of infinite heavy hole mass, whereas
it is lower in the energy when the finite value of the heavy hole mass is
taken into account. This reflects in the behavior of the effective mass of
2D carriers for their low concentration.

Thus, the presented results show that the interband mixing causes the
effective mass of 2D states to be larger than that calculated in the
infinite heavy hole mass approximation. This is more evident for $k^{+}$%
-``spin'' states and for large concentration of 2D carriers. The effective
mass of 2D carriers calculated with the most-used set of $\gamma $%
-parameters \cite{Guldner} differs mostly from the results of calculation,
in which the interaction with remote bands is neglected. To our knowledge
there are no experimental data concerning the effective mass of 2D states in
highly doped narrow gap semiconductors in case, when 2D states are in
resonance with the heavy hole valence band.

\section{Conclusion}

\label{sec:conc} The effects of interband mixing between 2D states and heavy
hole valence band states are found to influence the energy spectrum of 2D
states localized in a surface quantum well in the narrow gap semiconductor
with high doping level. They manifest themselves in the broadening of 2D energy
levels, which is sufficiently large for the energies close to the bottom of
2D subband, and the shift of 2D subbands to lower energies. The latter does not
practically affect the concentration dependence of subband distribution of
2D carriers, but leads to a change of their effective mass for high total
concentrations of 2D carriers.

\begin{figure}[tbp]
\caption{(a) Model of an insulator -- narrow gap semiconductor structure
with a surface quantum well used in the calculations. (b) The dispersion law 
$E(k,k_z=0)$ for carriers in the volume of semiconductors (solid curves).
The hatched regions show the continuous spectrum. The numbers I, II, and III
denote three different regions of the energy spectrum of 2D states (for
details, see text)}
\label{fig1}
\end{figure}

\begin{figure}[tbp]
\caption{The energy spectrum of 2D states calculated with different sets of $%
\gamma $-parameters: $\gamma _1=4.5$, $\gamma =1.0$ (the solid curves) and $%
\gamma _1=2.0$, $\gamma =0$ (the dashed curves), as compared with the energy
spectrum of 2D states calculated in the infinite heavy hole mass
approximation (the dotted curves). The value of the surface potential used
in the calculations is $285$ mV. $k^{+}$ and $k^{-}$ denote two different
``spin'' branches of the energy spectrum. The dot-dash curves show the
borders of region II, where the 2D states are in resonance with the heavy
hole valence band.}
\label{fig2}
\end{figure}

\begin{figure}[tbp]
\caption{The dependence of the electron concentration in each 2D subband on
the total concentration of 2D electrons. The notations and curve types are
the same as in Fig. 2.}
\label{fig3}
\end{figure}

\begin{figure}[tbp]
\caption{The dependence of the electron effective mass on the total
concentration of 2D electrons. The notations and curve types are the same
as in Fig. 2.}
\label{fig4}
\end{figure}


\begin{references}
\bibitem{Altarelli}  M.Altarelli, Phys. \ Rev. \ B\ {\bf 28}, 842 (1983).

\bibitem{Fassolino}  A.Fassolino and M.Altarelli, Surf. \ Sci. \ {\bf 142},
322 (1984).

\bibitem{Sakaki}  H.Sakaki, L.L.Chang, G.A.Sai-Halasz, C.A.Chang, and
L.Esaki, Solid \ State \ Commun. \ {\bf 26}, 589 (1978).

\bibitem{Sobkowicz}  Pawel Sobkowicz, Semicond. \ Sci. \ Technol. \ {\bf 5},
183 (1990).

\bibitem{Nachev}  I.Nachev, Nuovo \ Cimento {\bf 12 D}, 1143 (1990).

\bibitem{Ziegler}  A.Ziegler and U.R\"{o}ssler, Europhys. \ Lett. \ {\bf 8},
543 (1989).

\bibitem{Freytag}  B.Freytag, U.R\"{o}ssler, and O.Pankratov, Semicond. \
Sci. \ Technol. \ {\bf 8}, S243 (1993).

\bibitem{Kunze}  U. Kunze, Z.Phys. B -- Condensed Matter {\bf 80}, 47 (1990).

\bibitem{Kunze1}  J. M\"{u}ller and U. Kunze, Semicond. \ Sci. \ Technol. \ 
{\bf 8}, 703 (1990).

\bibitem{Bir-Pikus}  G.L.Bir and G.E.Pikus, {\em Symmetry and Strain Induced
Effects in Semiconductors} (Wiley, New York, 1974).

\bibitem{our}  G. M. Minkov, A. V. Germanenko, V. A. Larionova, and O. E.
Rut, Phys. \ Rev. \ B\ {\bf 54}, 1841 (1996).

\bibitem{Guldner}  Y.Guldner, C.Rigaux, A.Mycielski, and Y.Couder, Phys. \
Stat. \ Sol.\ b\ {\bf 82}, 149 (1977).

\bibitem{Marques}  G.E.Marques and L.J.Sham, Surf. \ Sci. \ {\bf 113}, 131
(1982).

\bibitem{ng}  G. M. Minkov, O. E. Rut, and A. V. Germanenko, (unpublished).

\bibitem{Bychkov}  Y.A Bychkov and E.I.Rashba, J. \ Phys. \ C \ {\bf 17},
6039 (1984).

\bibitem{Ohkawa}  F.J.Ohkawa and Y.Uemura, J. \ Phys. \ Soc. \ Jpn. \ {\bf 37%
}, 1325 (1974).

\bibitem{Wollrab}  R.Wollrab, R.Sizmann, F.Koch, J.Ziegler, and H.Maier,
Semicond. \ Sci. \ Technol. \ {\bf 4}, 491 (1989).

\bibitem{Sizmann}  R.Sizmann and F.Koch, Semicond. \ Sci. \ Technol. \ {\bf 5%
}, S115 (1990).

\bibitem{rus}  G. M. Minkov, O. E. Rut, V. A. Larionova, and A. V.
Germanenko, Zh. \ \'{E}ksp. \ Teor.\ Fiz. \ {\bf 105}, 719 (1994) [JETP {\bf %
78}, 384 (1994)].
\end{references}
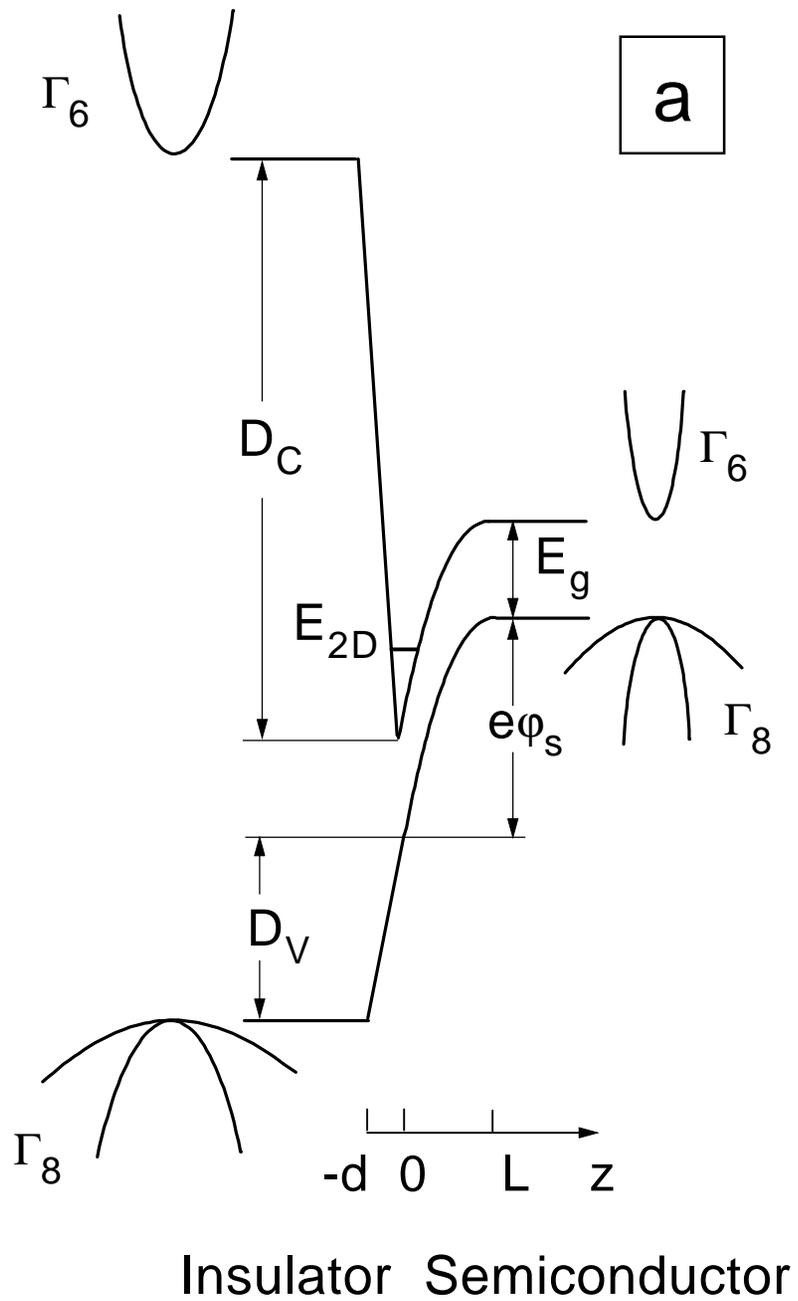
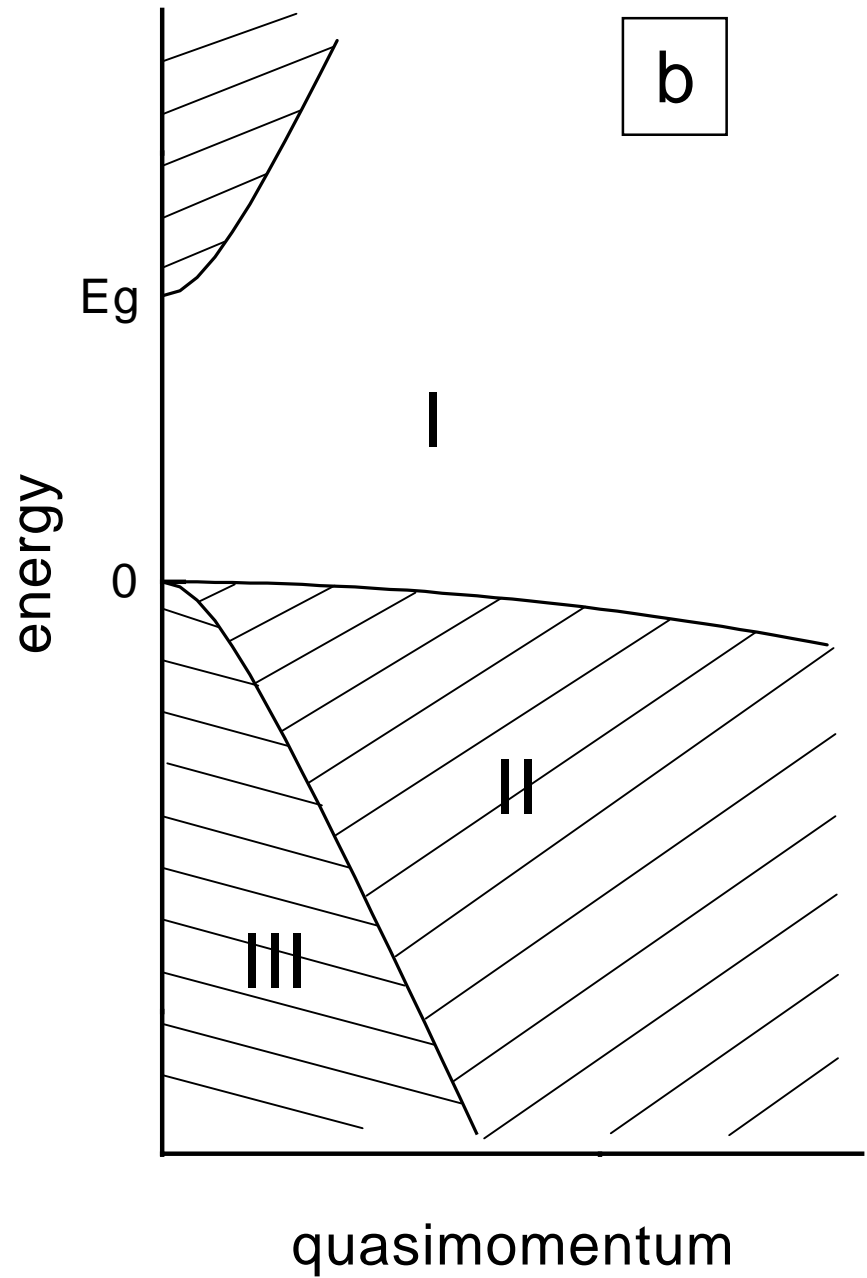
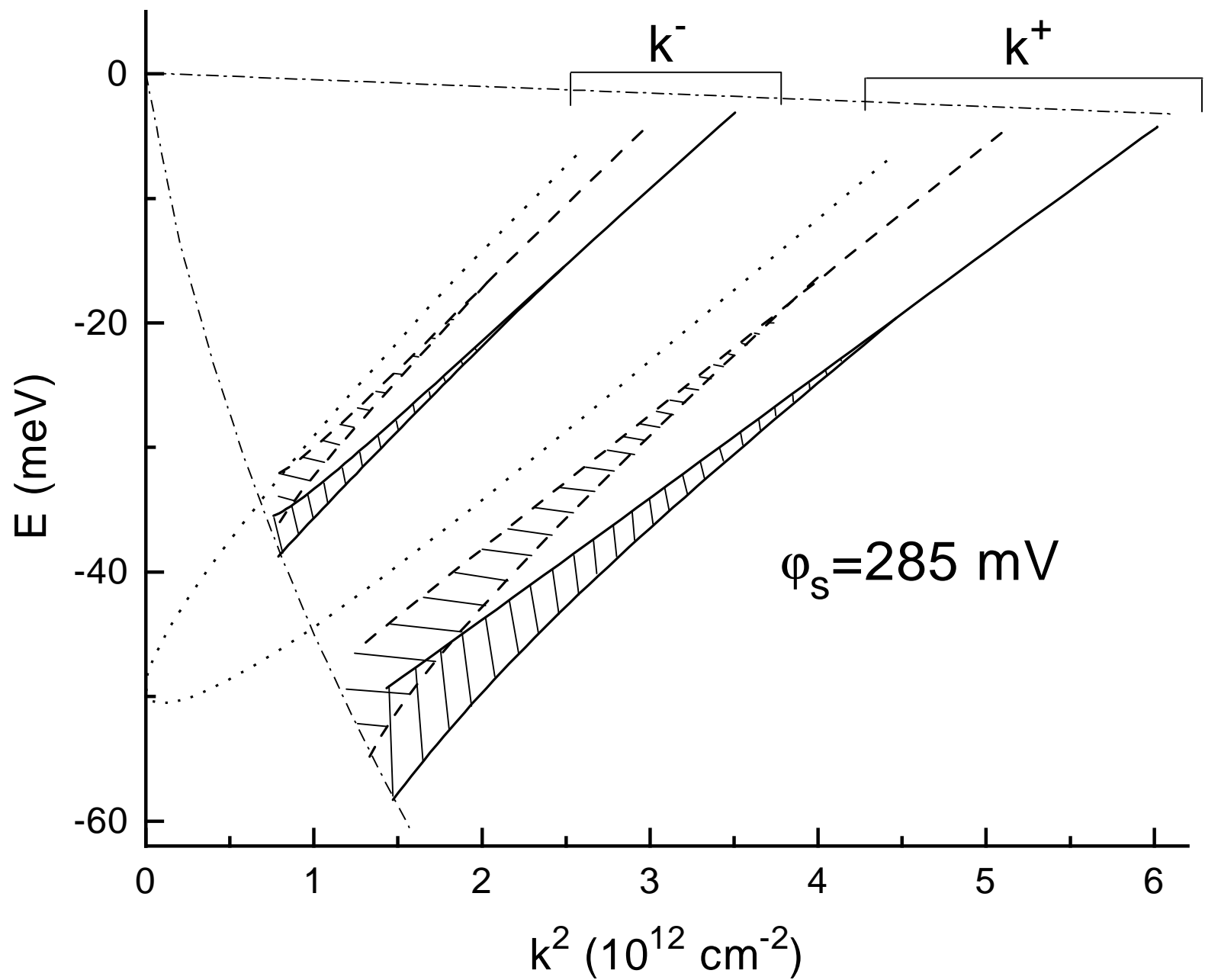
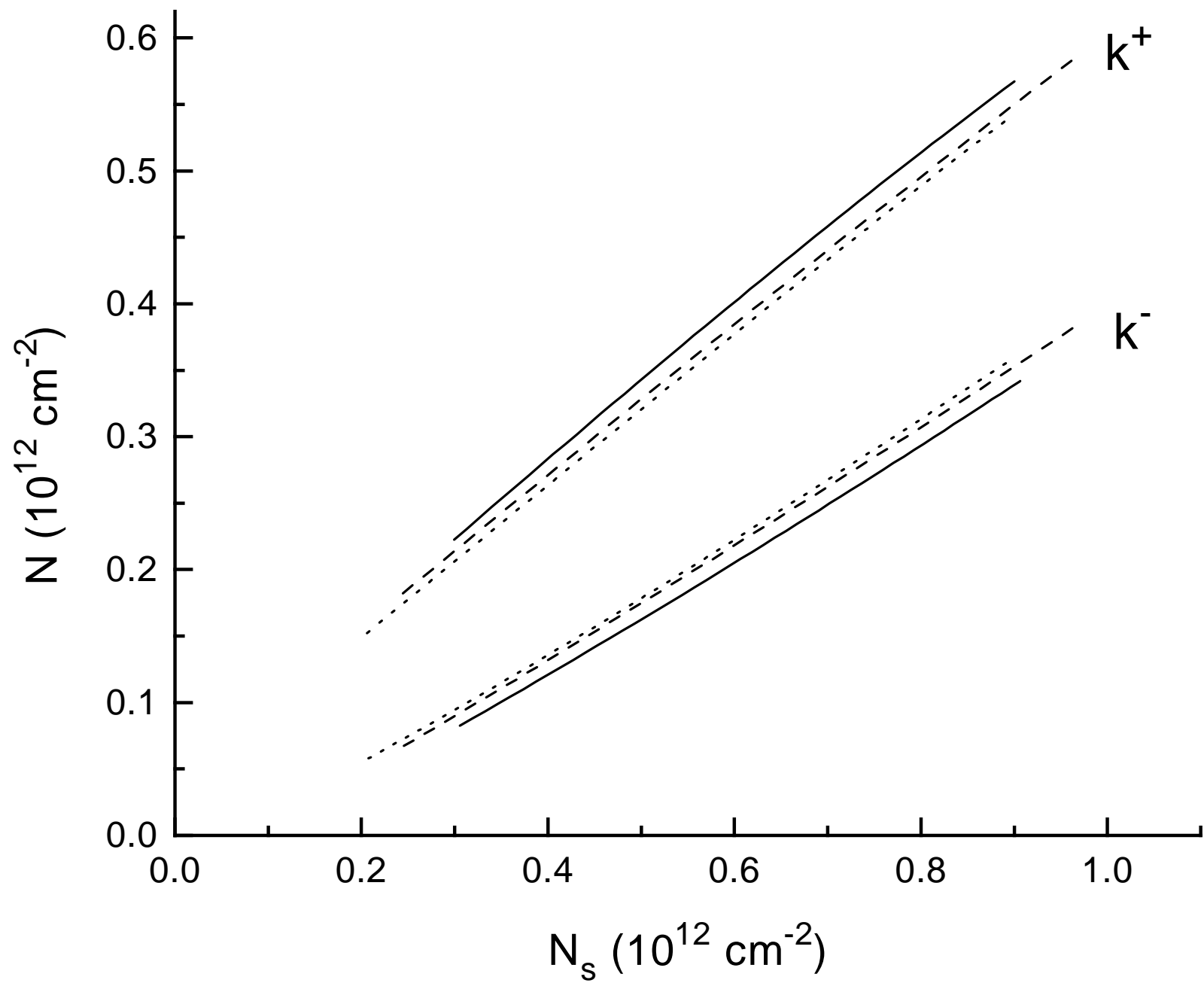
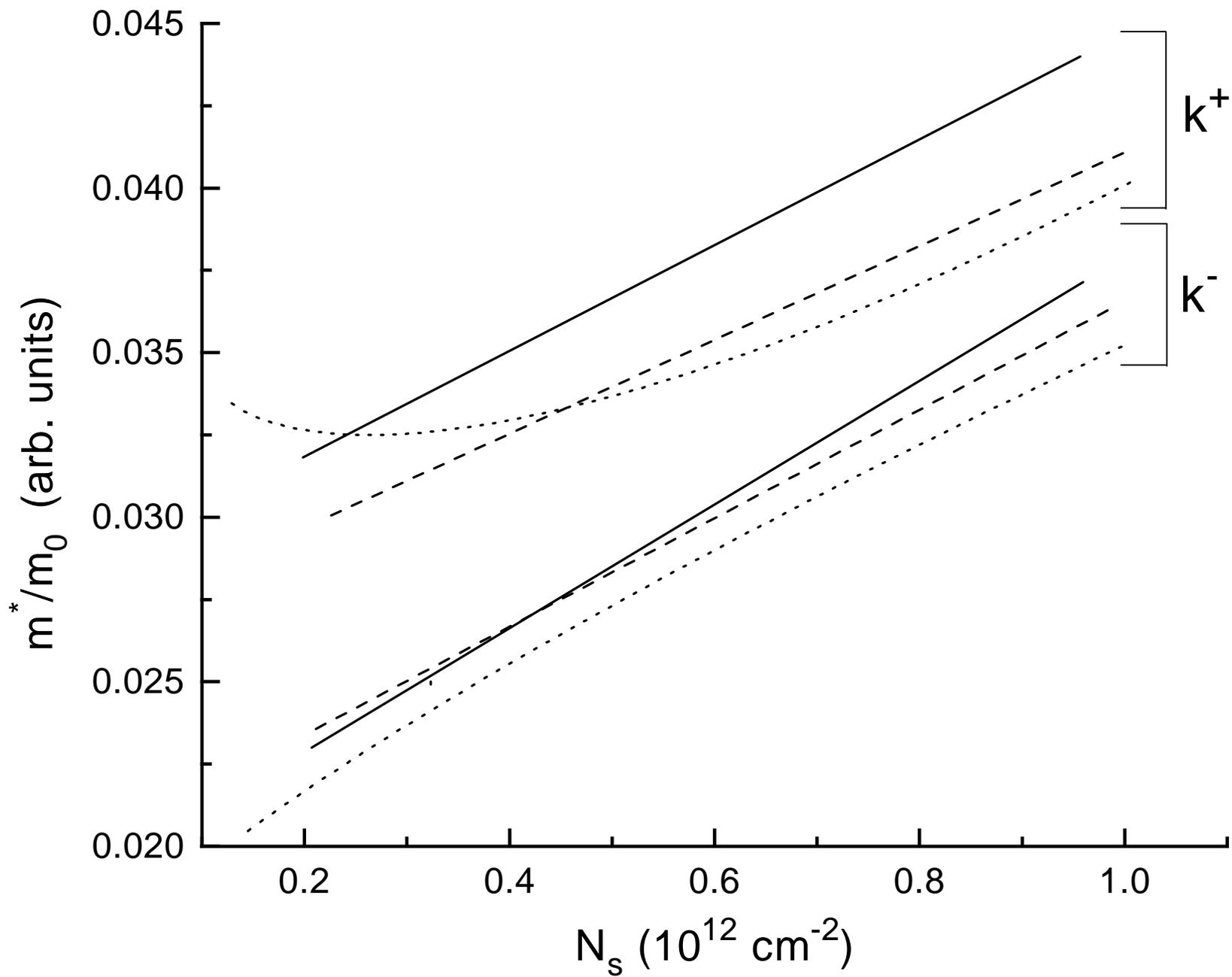
\end{document}